\documentclass[aps,preprint,amsmath,amssymb,amsfonts,nofootinbib]{revtex4-1}
\usepackage{epsfig}
\usepackage{graphicx}
\usepackage{dcolumn}
\usepackage{bm}
\usepackage{amsthm}
\usepackage{amsmath}
\usepackage{color}
\newcommand{\beq}{\begin{equation}}
\newcommand{\eeq}{\end{equation}}
\newcommand{\bea}{\begin{eqnarray}}
\newcommand{\eea}{\end{eqnarray}}

\begin{document}

\title{Spontaneous Symmetry Breaking, Conformal Anomaly and Incompressible Fluid Turbulence}
\author{Yaron Oz}
\affiliation{Raymond and Beverly Sackler School of Physics and Astronomy, Tel Aviv University, Tel Aviv 69978, Israel}

\date{\today}
\begin{abstract}
We propose an effective conformal field theory (CFT) description of steady state incompressible fluid turbulence at the inertial range of scales
in any number
of spatial dimensions. We derive
a KPZ-type equation for the anomalous scaling of the longitudinal velocity structure
functions and relate the intermittency parameter to the boundary Euler (A-type) conformal anomaly coefficient.
The proposed theory 
consists of a mean field CFT that exhibits Kolmogorov linear scaling (K41 theory) coupled
to a dilaton. The dilaton  is a Nambu-Goldstone gapless mode that arises from a spontaneous breaking 
due to the energy flux  of the separate scale and time symmetries of the inviscid Navier-Stokes equations to a K41 scaling with a dynamical exponent $z=\frac{2}{3}$.
The dilaton acts as a random measure that dresses the K41 theory and introduces intermittency.
We discuss the two, three and large number of space dimensions cases and how entanglement entropy can be used to characterize the intermittency strength.

\end{abstract}

%\pacs{...}

\maketitle

\section{Introduction}

Fully developed incompressible fluid turbulence is largely considered as the most important unsolved problem of classical physics.
One defines the inertial range to be the range of length scales $l \ll r \ll L$,  where the scales $l$ and $L$ are
determined by the viscosity and forcing, respectively. Experimental and numerical data suggest that
turbulence at the inertial range of scales  reaches a steady state that exhibits statistical homogeneity and isotropy
and is characterized by universal scaling exponents that depend only on the number of space dimensions $d$.

In \cite{Eling:2015mxa} we proposed an exact formula for the inertial range anomalous scalings $\xi_n$ of 
the  longitudinal structure functions $S_n$ :
\begin{equation}
S_n(r) = \langle (\delta v(r))^n \rangle  \sim r^{\xi_n}  \ , \label{structure}
\end{equation}
where
\begin{equation}
\xi_n -\frac{n}{3}= {\cal G}^2(d) \xi_n(1-\xi_n) \ , 
\label{KPZ}
\end{equation}
and  ${\cal G}(d)$ is a numerical real parameter that depends on the number of space dimensions $d \geq 2$.
It quantifies intermittency and the deviation from
Kolomogorov linear scaling $\xi_n =\frac{n}{3}$ \cite{Kolmogorov}.
$\delta v(r)$ is the longitudinal velocity difference between points separated by a fixed distance
$r=|\vec{r}|$,
$\delta v(r) = \left(\vec{v}(\vec{r},t) - \vec{v}(0,t)\right)\cdot \frac{\vec{r}}{r}$.
By $\langle \rangle$ we mean both integration over space (normalized by the volume of space) and 
over the probability distribution function of a  random force $F$, with $v^i=v^i(F)$. 
Note, that the dimensions $\xi_n$ in (\ref{KPZ}) are in units of length.

At leading order in the intermittency parameter ${\cal G}$, (\ref{KPZ}) coincides  with the  Kolmogorov-Obukhov model 
\cite{K62,Obukhov} :
\begin{align}
\xi_n - \frac{n}{3}  = {\cal G}^2 \frac{n}{3} \left(1-\frac{n}{3} \right). \label{KO}
\end{align}
They differ at high intermittency and  (\ref{KPZ}) 
may be viewed as a completion of the Kolmogorov-Obukhov formula to the strong intermittency regime.
Unlike (\ref{KO}) which implies physically inconsistent supersonic velocities at large $n$ and a violation
of the convexity inequality,  
its completion (\ref{KPZ}) is analytically and physically consistent \cite{Eling:2015mxa}. 
Formula (\ref{KPZ}) predicts a generic behaviour at finite $d$ (finite intermittency) and large $n$:  $\xi_n \sim \sqrt{n}$.

Formula (\ref{KPZ}) is consistent with the available numerical and experimental data  \cite{Eling:2015mxa} , however, it is a conjecture for
which we still lack an analytical derivation.
It is KPZ (Knizhnik-Polyakov-Zamolodchikov)-type relation \cite{Knizhnik:1988ak,Distler:1988jt} that arises
when coupling a dynamical system to a random geometry \cite{multi}. The inspiration for the proposal was the relation between fluids
and geometry, and hence between the statistics of fluid flows and the statistics of geometries (for a brief review 
see \cite{Eling:2010vr}).
By coupling to a random geometry one means changing the Euclidean measure $d \mu$ on  $R^d$ to a random measure 
\begin{equation}
d\mu_{\gamma}(x) = e^{\gamma\phi(x) -\frac{\gamma^2}{2}} d \mu \ , \label{random}
\end{equation}
where the Gaussian random field $\phi(x)$ has covariance $\phi(x)\phi(y) \sim - \log|x-y|$ when $l \ll |x-y| \ll L$ is in the inertial range. The intermittency parameter ${\cal G}$ is related to $\gamma$.
Intermittent features appear at short length scales, and this is when the effects of the random field $\phi$ are prominent.

The goal of this work is to propose an effective field theory framework of steady state incompressible fluid turbulence at the inertial range of scales, and derive  (\ref{KPZ}).
We will consider the energy cascade and work in the infinite Reynolds number limit.
The basic questions that arise are what is the definition of the random field $\phi(x)$ in the turbulent fluid field theory,
what is the field theory meaning of ${\cal G}(d)$ and how can we calculate it. 
We propose to identify the scalar field $\phi$ as a dilaton,
the Nambu-Goldstone (NG) gapless mode that arises from a spontaneous breaking 
due to the energy flux  of the separate scale and time symmetries of the inviscid Navier-Stokes (NS) equations to a K41 scaling with a dynamical exponent $z=\frac{2}{3}$.

 We define
the effective field theory on a $d$-dimensional closed ball in Euclidean space $R^d$ with radius $L$, $B_d(L)$, whose boundary is
the $(d-1)$-sphere $S^{d-1}$ of radius $L$. $L$ is the forcing IR scale when $d > 2$ \footnote{The two-dimensional case is special since the energy cascade is inverse: the forcing scale is $l$ and the dissipation scale is $L$.}.
The proposed effective field theory framework for the steady state dynamics of the turbulent fluid
is  conformally invariant. It 
consists of a mean field CFT that exhibits Kolmogorov linear scaling (K41 theory) coupled
to the dilaton. 
The odd $n$ structure functions $S_n$ (\ref{structure}) are non-vanishing  in K41 theory hence it is not Gaussian.
It is also non-local.
The dilaton acts as a random measure that dresses the K41 theory and introduces intermittency that leads 
to (\ref{KPZ}). 
This is reminiscent of two-dimensional
quantum gravity \cite{Knizhnik:1988ak,Distler:1988jt},  where matter  (our K41 fields) is coupled to a Liouville field (our dilaton) .
Here, however, both the K41 fields and the dilaton are made of the same hydrodynamic variables, the fluid velocity vector $v^i$
and the fluid pressure $p$.
We express the intermittency parameter ${\cal G}(d)$ in terms of 
the boundary Euler A-type conformal anomaly coefficient of the turbulent field theory and 
the number of space dimensions.
 We consider both odd and even number of space dimensions.

The paper is organized as follows.
In section 2 we analyze the spontaneous symmetry breaking of the separate space and time scale symmetries
of the inviscid NS equations due to the energy flux.
The breaking leads to the K41 scaling theory with a dynamical exponent $z=\frac{2}{3}$ and a gapless NG
dilaton mode. We construct the dilaton effective action which is scale invariant in the inertial range, and calculate the  dimensions
of scaling operators. 
In section 3 we propose the turbulent field theory and derive the KPZ scaling. We express the intermittency parameter
as a function of the Euler boundary conformal anomaly and the number of space dimensions 
(Eq. (\ref{branches}),(\ref{KPZfinal}),(\ref{intermittency}),(\ref{branches2})).
We propose the entanglement entropy as yet another tool to characterize the intermittency strength.
In section 4 we consider the physical cases of two and three space dimensions and the large $d$ limit.
Section 5 is devoted to a summary and outlook.
In the appendix we specify the notations and discuss the GJMS operator, the $Q$-curvature and the structure of the higher-dimensional Liouville field theory.

\section{Spontaneous Symmetry Breaking}

\subsection{Turbulent Flux States}

The incompressible NS equations provide a universal description of fluid flows at low Mach number, that is 
$v\ll v_s$ where $v_s$ is the speed of sound.  They read :
\begin{equation}
\partial_t v^i + v^j\partial_j v^i = -\partial_i p + \nu \partial_{jj} v^i + F^i, ~~~~~~\partial_iv^i = 0,~~~ i=1,...,d  \ ,
\label{NS}
\end{equation}
where $d$ is the number of space dimensions, $v^i, i=1...,d$  is the fluid velocity vector, $p$ is the fluid pressure, $\nu$ is the kinematic viscosity and $F^i$ is an external
random force.

Multiplying the first equation by $v^i$, summing over $i$, integrating over the ball $B_d(L)$ and using the second equation one gets 
\begin{equation}
\partial_t\int_{B_d} d^d x\left(\frac{v^2}{2}\right)  = \int_{B_d} d^d x \left(F^i v^i\right)   - \frac{\nu}{2} \int_{B_d} d^d x  \left(\partial_i v^j +  \partial_j v^i \right)^2  \ .
\label{Flux}
\end{equation}
The change in the kinetic energy of the fluid (\ref{Flux}) is dictated by the incoming energy flux due to the external force and by the energy 
dissipation due to viscosity.
In deriving (\ref{Flux}) we need to impose the boundary condition 
\begin{equation}
\int_{\partial B_d} \left(p+\frac{v^2}{2}\right) v^i d \Sigma_i  = 0 \ , \label{bou}
\end{equation}
where $d \Sigma_i$ is the directed surface measure of the spherical boundary $\partial B_d = S^{d-1}$. 
In a steady state we have the relation :
\begin{equation}
\int_{B_d} d^d x \left(F^i v^i\right)  =  \frac{\nu}{2} \int_{B_d} d^d x  \left(\partial_i v^j +  \partial_j v^i \right)^2 \ .
\label{steady}
\end{equation}

At very large Reynolds number  ${\cal R}_e = \frac{l_c v}{\nu}$, where $l_c$
is a characteristic length scale and $v$ the velocity difference at that scale, the viscosity term is very small compared
to the nonlinear term  $v^j\partial_j v^i$  in the NS equation. However the local energy dissipation $\epsilon(x) =  \frac{\nu}{2}  \left(\partial_i v^j +  \partial_j v^i \right)^2$ is nonvanishing
even in the limit  $\nu\rightarrow 0$ since the gradients of the velocity field
are singular. This is called the dissipative anomaly \cite{Frisch}.  
The steady state of turbulence is far from equilibrium since there is always a flow of energy flux and cascade, hence the Gibbs
measure is inappropriate for quantifying its statistics.

\subsection{Scale Symmetry Breaking}

In the absence of a viscosity term, the (inviscid) NS equations (\ref{NS}) exhibit two scale symmetries. These are independent scalings
of space and time which we denote by $R_x\times R_t$,
\begin{equation}
R_x : x^i \rightarrow e^{\sigma_1}x^i \ ,~~~~~~
R_t  : t \rightarrow e^{\sigma_2} t \ , \label{1}
\end{equation}
where an appropriate charge is assigned to the random force.
We denote the charges of any object under the two symmetries by $(r_x,r_t)$.  The charges of $v^i$ are $(1,-1)$,
and the charges of the different term in the inviscid NS equations are $(1,-2)$.
An alternative way of describing the two scale symmetries is
\begin{equation}
x^i \rightarrow e^{\sigma}x^i \ , ~~~~~ t \rightarrow e^{ z \sigma} t \ , \label{2}
\end{equation}
where $z$ is an arbitrary real parameter, and in relation to (\ref{1}):
$\sigma_1 = \sigma, \sigma_2 = z \sigma$.
For any value of $z$, one can define a total dimension of an object as 
\begin{equation}
\Delta_z = r_x + z r_t \ . \label{Deltaz}
\end{equation}
The dimension of $v^i$ is $\Delta_z=1-z$ and the dimension of the  different terms  in the inviscid NS equations is $\Delta_z = 1-2z$.

The symmetries (\ref{2}) can be broken from arbitrary $z$ to a particular value $z=z^{*}$.
We will denote this breaking by $R_x\times R_t \rightarrow R_{z^{*}}$.
The flux $F^iv^i$ at the boundary of the theory breaks spontaneously the symmetries  of the inviscid NS equations to $R_{z=\frac{2}{3}}$, as follows 
from the requirement that $\Delta_z(F^iv^i) = 2-3z = 0$.
This is the Kolmogorov scaling $\Delta_{K41}[v^i] = \frac{1}{3}$. Thus, we interpret the turbulent flux state
satisfying (\ref{steady}) as a state that spontaneously breaks (\ref{2}) to $z=\frac{2}{3}$.
One can instead consider the local energy dissipation $\epsilon(x)$ as the object whose expectation value breaks spontaneosuly
the scale symmetry. Note, that the Kolmogorov scaling is the critical scaling above which there is no dissipative anomaly
\cite{onsager}.

\subsection{The Dilaton Effective Action}

When a scale symmetry is spontaneously broken one expects a Nambu-Goldston gapless mode called the dilaton.
We denote the expectation value  that breaks $R_x\times R_t \rightarrow R_{\frac{2}{3}}$ by
$\langle F^iv^i \rangle = \Lambda$, where by $\langle \rangle$ we mean both integration over space (normalized by the volume of space) and 
over the probability distribution function of $F$ with $v^i=v^i(F)$.  The dilaton $\tau$ 
is the fluctuation seen by replacing the VEV :
\begin{equation}
\Lambda  \rightarrow \Lambda e^{ \delta \tau} \ , \label{mode}
\end{equation}
where $\delta$ is a c-number, and it is charged under  (\ref{1}) .

The dilaton effective action has to be invariant 
under Galilean boosts :
\begin{equation}
t\rightarrow t,~~~~ x^i \rightarrow x^i + v^i t,~~~~
\partial_i \rightarrow \partial_i,~~~~\partial_t \rightarrow \partial_t - v^i\partial_i,~~~~\tau \rightarrow \tau \ .
\label{boost}
\end{equation} 
This forbids time derivative terms in the Lagrangian and allows only space derivative ones 
(see \cite{Arav:2017plg} for a similar situation).
The dilaton effective action should also respect the symmetries (\ref{1}) with a dilaton invariant under the unbroken scale symmetry $R_{z=\frac{2}{3}}$
\cite{Hason:2017lao}.
We are interested in the steady state statistics and equal time correlation functions, hence we
can average over time $\frac{1}{T}\int_{0}^{T} dt $.

We will separate the discussion to two cases: an even number of space dimensions and an odd one.

\subsubsection{Even Number of Space Dimensions}

The dilaton effective action on $B_d(L)$ can be written (with arbitrary dimensionless coefficients) as  :
\begin{equation}
S_{dilaton} = S_0 + S_{Euler} + S_{CF} \ .
\end{equation}
$S_0$ reads :
\begin{equation}
{\cal S}_{0}=  - \int_{B_d} \sqrt{g} d^d x~ \tau\Box^{\frac{d}{2}}\tau   \ , \label{effdilaton}
\end{equation}
and it is the only local action in flat space that respects the symmetries (\ref{1}) with a dilaton invariant under the unbroken scale symmetry \cite{Hason:2017lao}.
It describes a log-correlated free scalar field that we will propose in the following to associate with the random scalar field in the random measure (\ref{random}).

${\cal S}_{Euler}$ reads :
\begin{equation}
{\cal S}_{Euler}=    -  \int_{\partial B_d} \sqrt{g_b} d^{d-1} x~  \tau q_d  \ , \label{back}
\end{equation}
where $\partial B_d = S^{d-1}$ is the spherical boundary of the ball, $g_b$ is the induced metric on the boundary and
$q_d$ is a boundary Euler density term constructed from the boundary intrinsic and extrinsic curvatures (\ref{Eback}).
The bulk Euler density $E_d$ is zero in (\ref{back}) since we consider a ball in Euclidean space.
The normalization  of $q_d$ is fixed in (\ref{chiBa}).

$S_{CF}$ consists of non Weyl invariant terms that can be written on the $S^{d-1}$ boundary
and take the general form :
\begin{equation}
S_{CF} = \int_{\partial B_d} \sqrt{g_b} d^{d-1} x~  H^{\mu_1\mu_2...}\partial_{\mu_1}
\tau \partial_{\mu_2}\tau... \ ,  \label{cf}
\end{equation}
where $H^{\mu_1\mu_2...}$ are made of the intrinsic and extrinsic curvatures
of the boundary.
The terms in $S_{CF}$ are suppressed by inverse powers of $L$, and since we work in the inertial range of scales 
 $r \ll L$ we will neglect them in the rest of the discussion. 
 
Thus,  the dilaton effective action consists of 
 (\ref{effdilaton}) and  (\ref{back}) 
with arbitrary dimensionless coefficients $ \kappa, Q$ :
\begin{equation}
S_{dilaton}=  - \kappa \int_{B_d} \sqrt{g} d^d x \tau\Box^{\frac{d}{2}}\tau  
+ Q   \int_{\partial B_d} \sqrt{g_d} d^{d-1} x~  \tau q_d  \ . \label{dilatonact}
\end{equation}
The action (\ref{dilatonact}) defines a higher derivative  nonunitary (non reflection positive) scale invariant field theory.
The Euler term in (\ref{dilatonact}) leads to an "anomalous"  non-conservation
of the current $J^{\mu} = 2 \partial^{\mu} \Box^{\frac{d-2}{2}} \tau$.
It has an interpretation of a background charge :
If we shift $\tau\rightarrow \tau + \sigma$ where $\sigma$ is a constant we pick up a charge ${\cal Q}$ from 
$e^{-S_{dilaton}}$,
$e^{{\cal Q} \sigma}$
where
\begin{equation}
{\cal Q} =   \alpha_d  Q  \ , \label{cQB}
\end{equation}
with
\begin{equation}
\alpha_d =   (4 \pi)^{\frac{d}{2}} (\frac{d}{2})!  \ , \label{alpha}
\end{equation}
and we used  (\ref{chiBa}) and (\ref{O}).

The background charge imposes a selection rule in correlation
functions and affects the  dimension of scaling operators as we will now calculate.

The two-point correlation function of $\tau$ reads
\begin{equation}
\langle \tau(\vec{x}_1)  \tau(\vec{x}_2)  \rangle = \frac{d}{2 \alpha_d\kappa}\left(\log\frac{L}{|\vec{x}_{12}|} \right) + ... \ . \label{twopoint}
\end{equation}
$\vec{x}_{12} = \vec{x}_1 - \vec{x}_2$, $L$ is the IR regulator  and dots refer to a constant and additional terms in (\ref{twopoint}) that vanish when $\L\rightarrow \infty$. 
$\tau$ is not a scaling operator and one should consider operators
that transform covariantly under scaling.
Consider the correlation function of the operators $e^{\beta_i\tau}$. The background charge imposes a selection rule :
\begin{equation}
\sum_i \beta_i + {\cal Q} = 0 \ . \label{selection}
\end{equation}
The two point function is :
\begin{equation}
\langle \exp^{\beta\tau(\vec{x}_1)}  \exp^{-(\beta+{\cal Q})\tau(\vec{x}_2)}  \rangle = |\vec{x}_{12}|^{-2 \Delta} \ , \label{sc}
\end{equation}
where
\begin{equation}
\Delta = - \frac{d}{4 \alpha_d \kappa} \beta (\beta + {\cal Q})  \ , \label{Delta}
\end{equation}
is the scaling dimension in units of $(length)^{-1}$.

\subsubsection{Odd Number of Space Dimensions}

The case with odd number of space dimensions is more subtle. There are no local bulk action terms that respect the symmetries, and
there is a fractional derivative in $S_0$ (\ref{effdilaton}) which makes 
the Lagrangian  nonlocal in space. 
The Euler term takes the form :
\begin{equation}
{\cal S}_{Euler} =   \int_{\partial B_d} \sqrt{g_b}  d^{d-1} x~  \tau E_{d-1}  \ , \label{back1}
\end{equation}
where $E_{d-1}$ is the Euler density of the boundary $S^{d-1}$ sphere normalized as in (\ref{normalization}).
There is an additional local term that can be written on the boundary while neglecting terms suppressed by inverse powers of $L$ :
\begin{equation}
{\cal S}_{Boundary} =    \int_{\partial B_d} \sqrt{g_b} d^{d-1} x~\tau\Box^{\frac{d-1}{2}}\tau  \ . \label{bound1}
\end{equation}
Thus, the dilation effective action includes the non-local term (\ref{effdilaton}), the Euler term (\ref{back1}) and the 
additional local boundary term (\ref{bound1}) with dimensionless couplings : 
\begin{equation}
S_{dilaton}=  - \kappa  \int_{B_d} \sqrt{g} d^d x~\tau\Box^{\frac{d}{2}}\tau - 
\tilde{\kappa} \int_{\partial B_d} \sqrt{g_b} d^{d-1} x~\tau\Box^{\frac{d-1}{2}}\tau   -  Q  \int_{\partial B_d} \sqrt{g_b}  d^{d-1} x~  \tau E_{d-1}    \ . \label{dilatonact2}
\end{equation}
Matching the solution (\ref{twopoint}) at the boundary as dictated by the first and second terms in (\ref{dilatonact2}) implies :
\begin{equation}
\tilde{\kappa} = \frac{(d-1)\alpha_d}{d \alpha_{d-1}} \kappa  \ . \label{Matching}
\end{equation}
Formulas (\ref{selection}), (\ref{sc}) and (\ref{Delta}) continue to hold with :
\begin{equation}
{\cal Q} =   2 \alpha_{d-1} Q  \ , \label{cQB2}
\end{equation}
and we used (\ref{normalization}) and (\ref{O}).

\subsubsection{Discussion}

There are a couple of remarks in order. On the one hand, the fact that the dilaton carries charge under the time scaling symmetry $R_t$ (\ref{1}) 
effectively forbids dilaton exponential terms  $e^{\alpha\tau}$  in the steady state dilaton effective action. On the other hand we seem to allow operators
of such structure and calculate their two-point function (\ref{sc}), which looks inconsistent. In fact, when we will construct the operators of the turbulence field theory  in the next section the dilaton exponential terms
will not stand by themselves. Rather, they will be used to dress
the K41 field theory operators such that the whole operator should be invariant under $R_t$.
In general, all the physical quantities that will appear later in calculations will be required to be invariant under $R_t$.

Physically, the K41 degrees of freedom and the dilaton are made of the same fluid variables: the velocity and the pressure. Seperating the discussion to first constructing the dilaton effective action in this section and then dressing the K41 operators in the next section is in some sense artificial and simply reflects our ignorance of the detailed combined dynamics until steady state is reached.

In the odd-dimensional case we included a non-local term in the effective action. This opens up a pandora box since there 
is an  infinite number of non-local terms that respect the symmetries and could have been included in the action. 
A proper physical justification for neglecting them
will probably follow from an understanding of the combined K41 fields and dilaton dynamics.
A mathematical explanation for neglecting them can come from a better understanding of conformally covariant
pseudo-differential operators (see appendix A).

\section{Turbulence Field Theory}

We would like to construct a $d$-dimensional effective field theory for equal time correlation functions of steady state incompressible fluid turbulence. The number of space dimensions $d$
can be even or odd.
The Kolomgorov linear scaling theory is a mean field theory (K41), and
we will denote its set of degrees of freedom by $\varphi$.
It is non-local and non-Gaussian field theory.
We propose to define the complete theory as a dressing of the K41 theory by the dilaton $\tau$.
The dilaton accounts for the fluctuations around the mean field  $\varphi$ that result in the intermittency.
This is reminiscent of two-dimensional
quantum gravity \cite{Knizhnik:1988ak,Distler:1988jt},  where matter  (our K41 fields) is coupled to a Liouville field (our dilaton).
Here, however, both $\varphi$ and $\tau$ are made of the same hydrodynamic variables, the velocity vector $v^i$
and the pressure $p$. Also, unlike  two-dimensional
quantum gravity, here diffeomorphisms are not gauge symmetries and therefore there are no gauge fixing ghosts.
The turbulence field theory is nonunitary
since conserved quantities leak from the flux states \cite{Polyakov:1992er}.

As to symmetries, the field theory should exhibit $d$-dimensional translations and rotations and spatial scale invariance $R_x$ (\ref{1}). We propose that it posseses  conformal invariance 
 (suggested for two dimensions in \cite{Polyakov:1992er}) \footnote{Note, that nonunitary scale invariant field theories are not  necessarily conformal invariant.
 An example is the field theory of elasticity \cite{Riva:2005gd}.}.

\subsection{KPZ Scaling and Critical Exponents}

We define the random metric $\bar{g}_{ij}$  by dressing the background metric  $g_{ij}$  with a dilaton factor :
\begin{equation}
\bar{g}_{ij} =  e^{2\gamma \tau} g_{ij},~~~~\sqrt{\bar{g}} = e^{d\gamma \tau}\sqrt{g} \ ,
\label{dress}
\end{equation}
and require that the operator $e^{d\gamma \tau}$ is a conformal operator of scaling dimension $d$. Using 
(\ref{Delta})
we have
\begin{equation}
-\frac{d\gamma (d\gamma + {\cal Q})}{4 \alpha_d \kappa} = 1 \ .   \label{Deltakpz}
\end{equation}
Thus,
\begin{equation}
{\cal Q} = -d \gamma - \frac{4 \alpha_d \kappa}{d\gamma}  \ ,   \label{Deltakpz1}
\end{equation}
and
\begin{equation}
\gamma_{\pm} = \frac{-{\cal Q} \pm \left({\cal Q}^2 - 16 \alpha_d \kappa  \right)^{\frac{1}{2}}}{2 d} \ . \label{plusminus}
\end{equation}

We construct the operators of the turbulence field theory $\hat{O}$ by dressing the K41 field theory operators $O$ with a dilaton factor :
\begin{equation}
\hat{O}(x) = e^{\beta \tau} O (x),~~~~~\beta= d \gamma(1-\Delta) \ . 
\label{dress2}
\end{equation}
As discussed in the previous section, we require that the  $R_t$  charge of the dilaton exponential in (\ref{dress2}) will
be cancelled by the that of the K41 operator $O$ making  $\hat{O}(x)$  $R_t$ invariant.
Consider the dressed operator $O(x)$ (\ref{dress2}),
and let $d\Delta_0$ denote the undressed dimension of $O$.
We require that  the scaling dimension of $\hat{O}(x)$ is $d$, thus :
\begin{equation}
- \frac{\beta(\beta + {\cal Q})}{4 \alpha_d\kappa }  + \Delta_0 =  1 \ . \label{bet}
\end{equation}
Solving (\ref{bet}) we have the KPZ-type  relation :
\begin{equation}
\Delta - \Delta_0  = \frac{d^2\gamma^2}{4 \alpha_d\kappa }\Delta(1-\Delta) \ . \label{dd0}
\end{equation}
for the dressed dimension $\Delta$. 
The quantity :
\begin{equation}
{\cal G}^2 =   \frac{d^2\gamma^2}{4 \alpha_d\kappa }   \label{intermittency1}
\end{equation}
  in (\ref{dd0})   is the intermittency parameter ${\cal G}^2(d)$ in (\ref{KPZ}). 

Consider now the partition function of the turbulent field theory as a function of a fixed random volume $V$:
\begin{equation}
Z(V) = \int D \varphi D \tau e^{-S} \delta\left(\int \sqrt{g}e^{d\gamma \tau} d^d x - V \right) \ , \label{Z}
\end{equation}
where $S$ is the turbulent field theory action that includes the K41 fields and the dilaton.
Using a similar argument to that in \cite{Distler:1988jt}, we shift $\tau \rightarrow \tau + \frac{\sigma}{d\gamma }$ where $\sigma$ is 
constant. This leads to the relation $Z(V) = e^{-\sigma(1- \frac{{\cal Q}}{d \gamma})} Z(e^{-\sigma} V)$, hence at large $L$
\begin{equation}
Z(V) \sim V^{-1 + \frac{{\cal Q}}{d \gamma}}  \ , \label{L}
\end{equation}
where we used (\ref{Deltakpz1}) and (\ref{intermittency1}).
The exponent 
is reminiscent of the string susceptibility and it would be interesting to find a way to check this experimentally and numerically.

Consider next the expectation value of the one-point function :
\begin{equation}
F_{O} (V) = Z(V)^{-1} \int D \varphi D \tau e^{-S} \delta\left(\int \sqrt{g}e^{d\gamma \tau} d^d x - V \right) 
\int \sqrt{g}  e^{d\gamma (1- \Delta) \tau} O d^d x 
\  . \label{F}
\end{equation}
Shifting by a constant as before we see that 
\begin{equation}
F_{O} (V) \sim V^{1-\Delta} \ , \label{Fdelta}
\end{equation}
where $\Delta$ is  the dressed dimension 
that satisfies the KPZ relation (\ref{dd0}).
%Note, that with our dressing conventions the dimension $\Delta$ is of opposite sign compared to the standard CFT 
%conventions (\ref{Delta}) and is compatible with (\ref{KPZ}).

%Note that the objects that appear in the calculations (\ref{Z}) and (\ref{Fdelta}): the background metric $\sqrt{\hat{g}}e^{d\gamma \tau}$ and 
%the dressed operators do not  carry $R_t$ charges.

\subsection{Conformal Anomaly}

In the following we will use the requirement of conformal invariance to fix the intermittency parameter in 
terms of the boundary Euler anomaly coefficients of the K41 and the dilaton field theories.

\subsubsection{Even Number of Space Dimensions}

Conformally  invariant relativistic field theories  in even number of dimensions  exhibit conformal  anomalies
\cite{Deser:1993yx}. This can be revealed in flat space correlation functions and via the one-point function
of the trace of the  stress-energy tensor in a bulk curved background or at the boundary.
Since our bulk space is flat the one-point function  bulk anomaly vanishes 
\begin{equation}
\langle T^{\mu}_{\mu} \rangle_{bulk} =  0 \ . 
\end{equation}
We do have an anomaly localized 
at the boundary $r=L$. The boundary is conformally flat hence the relevant anomaly is the Euler A-type
one : 
\begin{equation}
\langle T^{\mu}_{\mu} \rangle_{boundary} =   (-)^{\frac{d}{2}} a q_d   \ , \label{anomaly}
\end{equation}
where $a$ is the boundary anomaly c-number coefficient.

We assume that under a Weyl transformation of the metric $g_{ab}  \rightarrow e^{2\sigma} g_{ab}$ the K41 field theory action is invariant :
\begin{equation}
S_{K41}(e^{2\sigma} g, \varphi_{\sigma}) = S_{K41}(g, \varphi)  \ ,
\end{equation}
where by $ \varphi_{\sigma}$ we denote the Weyl transformation of $\varphi$.
$S_{K41}$ is a formal unknown expression that is expected to be non-local.
There is a quantum anomaly
due to the non-invariance of the measure of the form :
\begin{equation}
D_{e^{2\sigma} g} \varphi = e^{-a_{K41} S_l(\sigma)} D_{g} \varphi  \label{a-anomaly} \ ,
\end{equation}
where $S_l$   given by  (\ref{Liouvillef}) (see  \cite{Komargodski:2011vj,Elvang:2012st,Elvang:2012yc,Herzog:2015ioa})
\begin{equation}
S_l (\sigma)  = \left[ \int_{B_d} \sqrt{g} d^d x  \frac{d}{2} \sigma\Box^{\frac{d}{2}}\sigma    - (-1)^{\frac{d}{2}}  \int_{\partial B_d} \sqrt{g_d} d^{d-1} x~     \sigma q_d \right]  \ , \label{Liouville}
\end{equation}
is the $d$-dimensional Liouville action, and we neglected the terms that are suppressed by inverse powers of $L$.
 $a_{K41}$ is the Euler anomaly coefficient of the K41 conformal field theory.

Under a Weyl transformation the dilaton field theory action (\ref{dilatonact}) is not invariant :
\begin{equation}
S_{dilaton}(e^{2\sigma} g, \tau - \sigma) = S_{dilaton}(g, \tau)  -  S_{dilaton}(g, \sigma)  \ , \label{dilsigma}
\end{equation}
where to be consistent with the analysis below we used $Q = (-)^{\frac{d}{2}}\frac{2\kappa}{d}$. With
this choice, the dilaton action is (up to the overall coefficient $\kappa$) the higher-dimensional Liouville action (\ref{Liouville}), and (\ref{dilsigma}) follows from  (\ref{anomalytransformation}).
There is a quantum anomaly
due to the non-invariance of the measure :
\begin{equation}
D_{e^{2\sigma} g} \tau = e^{-a_{dilaton} S_l(\sigma)} D_{g} \tau \label{a2-anomaly} \ ,
\end{equation}
where $a_{dilaton}$ is the dilaton Euler anomaly coefficient.
We will require conformal invariance of the turbulence field theory, and hence a cancellation of the conformal anomaly :
\begin{equation}
(a_{K41} + a_{dilaton}) S_l(\sigma) - S_{dilaton}(\sigma)  = 0 \ . \label{cancel}
\end{equation}
Denote :
\begin{equation}
a_{+} = a_{K41} +  a_{dilaton},~~~~\bar{a}_{+} = \alpha_d a_{+}  \ ,
\end{equation}
with $\alpha_d$ in (\ref{alpha}).
The condition for the cancellation of the conformal anomaly (\ref{cancel}) gives :
\begin{equation}
\kappa =  -\frac{d}{2} a_+,~~~{\cal Q} = -(-)^{\frac{d}{2}}  \bar{a}_{+}  \ . \label{a}
\end{equation}
Using (\ref{plusminus}) we get :
\begin{equation}
\gamma_{\pm} =  \frac{ (-)^{\frac{d}{2}}  \bar{a}_{+} \pm \left(\bar{a}_{+}^2 + 8 d  \bar{a}_{+} \right)^{\frac{1}{2}}}{2 d} \ . \label{branches}
\end{equation}
The requirement  for real solution in (\ref{branches}) implies that 
\begin{equation}
\bar{a}_{+}  = \bar{a}_{crtitical} \leq - 8 d \ . \label{bd}
\end{equation}

The KPZ-type  relation (\ref{dd0}) reads :
\begin{equation}
\Delta - \Delta_0  = - \frac{d \gamma^2}{2 \bar{a}_{+}}  \Delta(1-\Delta)\ , \label{KPZfinal}
\end{equation}
for the dressed dimension $\Delta$. 
The  intermittency parameter (\ref{intermittency1})  is expressed in (\ref{KPZfinal}) as a function
of the number of space dimensions $d$ and the sum of the Euler anomaly coefficients of the K41 and dilaton field theories $a_+$ :
\begin{equation}
{\cal G}^2 =  - \frac{d \gamma^2}{2 \bar{a}_{+}}  \ . \label{intermittency}
\end{equation}
The condition (\ref{bd}) implies that ${\cal G}^2 \geq 0$.
We also have the relation :
\begin{equation}
\gamma_{+}\gamma_{-} = -\frac{2 \bar{a}_{+}}{d}  \ , \label{gammafinal}
\end{equation}
or alternatively, 
\begin{equation}
{\cal G}_-{\cal G}_+ = 1 \ . \label{one}
\end{equation}
For a fixed number of space dimensions and a given $a_+$ there are two branches in (\ref{branches}).
When $\frac{d}{2}$ is even:  (+) $1 \leq {\cal G}^2_+$, where ${\cal G}_+^2(\bar{a}_{critical}) = 1$ and  ${\cal G}^2_+(\bar{a}_+ \rightarrow \infty) = \infty$; (-)  $0 \leq {\cal G}^2_- \leq 1$,
 where ${\cal G}^2_-(\bar{a}_{critical}) = 1$ and   ${\cal G}^2_-(\bar{a}_+ \rightarrow  \infty) = 0$,
while  when $\frac{d}{2}$ is odd they are exchanged : ${\cal G}^2_+ \leftrightarrow {\cal G}^2_-$

\subsubsection{Odd Number of Space Dimensions}

In odd dimensions there is no bulk conformal anomaly but there is a boundary one \cite{Graham:1999pm}.
In our case the boundary is the $(d-1)$-sphere and relevant conformal anomaly is the boundary Euler anomaly :
\begin{equation}
\langle T^{\mu}_{\mu} \rangle_{boundary}   =   (-)^{\frac{d+1}{2}}a E_{d-1}  \ ,
\end{equation}
where $a$ is the boundary anomaly coefficient.
As in the even number of space dimensions case, we assume that under a Weyl transformation of the metric the K41 field theory action is invariant.
There is a quantum boundary conformal anomaly
due to the non-invariance of the measure of the form
\begin{equation}
D_{e^{2\sigma} g} \varphi = e^{-a_{K41} S_{l,boundary}(\sigma)} D_{g} \varphi  \label{aodd-anomaly} \ ,
\end{equation}
where $S_l$ is given by 
the boundary Liouville action (\ref{Liouvillef}) (see e.g. \cite{Jensen:2015swa}) is :
\begin{equation}
S_{l,boundary} =  \int_{S^{d-1}} \sqrt{g_b} d^{d-1} x \left(\frac{d-1}{2} \sigma\Box^{\frac{d-1}{2}}\sigma  +  (-)^{\frac{d-1}{2}}  \sigma E_{d-1}   \right)  \ , \label{Liouvilleb}
\end{equation}
and we neglected the terms that are suppressed by inverse powers of $L$.
 $a_{K41}$ is the Euler boundary anomaly coefficient of the K41 conformal field theory.

Under a Weyl transformation the dilaton field theory action is not invariant and one gets that its local part satisfies (\ref{dilsigma}).
There is also a quantum conformal boundary anomaly
due to the non-invariance of the measure
\begin{equation}
D_{e^{2\sigma} g} \tau = e^{-a_{dilaton} S_{l,boundary}(\sigma)} D_{g} \tau \label{a1-anomaly} \ ,
\end{equation}
where $a_{dilaton}$ is the dilaton Euler boundary anomaly.
We will require conformal invariance of the turbulence field theory on $B_d$, and hence a cancellation of the conformal anomaly 
(\ref{cancel}). 
Using (\ref{dilatonact2}) (\ref{Matching}) (\ref{cQB2}) and  (\ref{Liouvilleb})  we get
\begin{equation}
\tilde{\kappa} =  -\frac{d-1}{2} a_+,~~~\kappa =  -\frac{d}{2}\frac{\bar{a}_+}{\alpha_d} ,~~~ {\cal Q} =   (-)^{\frac{d+1}{2}}  2 \bar{a}_+  \ ,
\end{equation}
where
\begin{equation}
\bar{a}_{+} = \alpha_{d-1} a_{+}   \label{aodd} \ .
\end{equation}
Repeating the same steps as in the even-dimensional case we have :
\begin{equation}
\gamma_{\pm} =  \frac{ (-)^{\frac{d-1}{2}}  \bar{a}_{+} \pm \left(\bar{a}_{+}^2 + 2 d  \bar{a}_{+} \right)^{\frac{1}{2}}}{d} \ . \label{branches2}
\end{equation}
The requirement  for real solution in (\ref{branches2}) implies that 
\begin{equation}
\bar{a}_{+}  = \bar{a}_{crtitical} \leq - 2 d \ . \label{bdodd}
\end{equation}
The KPZ-type equation (\ref{gammafinal})  and (\ref{one}) continue to hold with the definition (\ref{aodd}).
The analysis of the $(\pm)$ branches is similar except  that one replaces the $\frac{d}{2}$ odd and even cases by 
 $\frac{d+1}{2}$ odd and even cases, respectively.

The cancellation (\ref{cancel}) is of the local terms and one remains with non-local terms that
violate conformal invariance. As discussed in the appendix, it is plausible that there is an analog of the action (\ref{actionQ})
for the bulk fractional derivative operator, which can be used to cancel the non-local terms. Another possibility
is that these non-local violating terms should be cancelled by the K41 field theory.
Clearly this issue needs further study.

\subsubsection{Summary}

We proposed to associate with  $(\delta v(r))^n$  in (\ref{structure}) a CFT operator
$\hat{O} = e^{-d\gamma (1- \Delta) \tau} O_{K41}$ (\ref{dress}), where  the K41 field theory operator $O_{K41}$ is 
dressed by the dilaton factor. The map between the fluid variables and the CFT variables is such that 
$\Delta(O_{K41}) = d \Delta_0 = \frac{dn}{3}$ and $\Delta = \xi_n$ is the anomalous scaling (\ref{structure}). 
More generally, 
we can consider not just the anomalous dimensions but also correlation functions $\langle \hat{O}_1(r_1)... \hat{O}_N(r_N) \rangle $
that may be checked numerically or experimentally. 
Note, that the boundary anomaly coefficient $a_+$ depends on the boundary condition of $\varphi$ and $\tau$ on $\partial B_d$.
This should probably be related to (\ref{bou}) and 
the forcing.

As noted, for a fixed number of space dimensions and a given $a_+$ there are two branches in (\ref{branches}) and (\ref{branches2}).
One may argue, that the "semi-classical" limit  of large $|a_+|$ corresponds to weak intermittency, and 
indicates the choice of the branch. This is compatible with the experimental and numerical data in three and four space
dimensions where the intermittency parameter ${\cal G}^2$  is smaller than one \cite{Eling:2015mxa}.
However, it is not clear that the intermittency parameter cannot exceed one in general, which requires the other choice of branch.

\subsection{The Conformal Anomaly Coefficients}

The Euler conformal anomaly coefficient of the dilaton or K41 CFT can be obtained from its partition function $Z_{B_d}$ on the ball $B_d(L)$ :
\begin{equation}
a \sim \frac{\partial}{\partial \log L} \log Z_{B_d}(L) = - \int \sqrt{g} d^d x~ \langle T^{\mu}_{\mu} \rangle \ . \label{Za}
\end{equation}
This calculation can be done for the dilaton field theory but not yet for the K41 CFT.

\subsubsection{The Dilaton Theory}

Consider an even number of space dimensions.
The are two terms in the dilaton partition function that contribute to the 
anomaly coefficient : One contribution is from $S_0$ (\ref{effdilaton}) and the second from
the background charge term  (\ref{back}) \footnote{I thank T. Levy for a discussion on this point.}.
Denote this by  $a_{dilaton} = a_0 + a_{charge}$. 

$S_0$ (\ref{effdilaton}) can be coupled to gravity in a  Weyl invariant way using the 
Weyl covariant GJMS operators \cite{GJMS}. On a constant curvature even-dimensional conformally flat manifold of dimension $d$
 the higher derivative GJMS operator factorizes as a product of Laplacians with masses. The conformal anomaly
coefficient can be calculated and one gets in our notation \cite{Dowker:2010bu} :
\begin{equation}
a_{0} = - \frac{1}{d!  \alpha_d} \int_0^{\frac{d}{2}}dt~ \prod_{i=0}^{\frac{d}{2}-1}\left(i^2-t^2\right)  \ . \label{adil}
\end{equation}
$a_0$ is positive when $\frac{d}{2}$ is odd and negative when it is even. 
Note, that if we integrate from zero to one in (\ref{adil}) we will get the anomaly coefficient of a free scalar in $d$ dimensions.

The contribution of the background charge to the partition function gives:
\begin{equation}
\log Z = - \frac{1}{2} \bar{a}_+ \log L \ ,
\end{equation}
where we used (\ref{twopoint}), (\ref{a}) and (\ref{chiBa}).
The background charge contribution to the dilaton anomaly coefficient is :
\begin{equation}
a_{charge} = -\frac{(-)^\frac{d}{2}}{2} a_+ \ . \label{backcharge}
\end{equation}
As we will see, the value of $a_0$ is negligible compared to the value needed for $a_{+}$ when we match
to the turbulence data, hence $a_{dilaton} \simeq a_{charge}$.
We are interested in the boundary anomaly and its precise value depends on the boundary conditions that will be imposed. 
However, this will not change significantly neither the values nor their large $d$ behaviour.
The same calculation can be done for an odd-dimensional bulk with an even-dimensional boundary where
$a_{charge} = \frac{(-)^\frac{d+1}{2}}{2} a_+$ and
the same conclusion holds.

\subsubsection{K41 Effective Field Theory}

The Kolomgorov linear scaling theory is a non-local theory, perhaps of vortices.
It assumes that the mean viscous energy dissipation rate $\epsilon$ is constant in the limit of infinite Reynolds number,
from which a linear scaling of the exponents follows.
In K41 theory the random velocity field is self-similar which misses the intermittency of the turbulent flows.
Although we know the spectrum of scaling dimensions, we need more information in order to 
construct the field theory and calculate the anomaly coefficient $a_{K41}$.

Let us first estimate the number of degrees of freedom of the K41 mean field theory following Landau's argument.
We assume that K41 is a theory of vortices of size $k_{\nu}^{-d}$, where 
$k_{\nu} \sim \left(\frac{\epsilon}{\nu^3}\right)^{\frac{1}{4}}$ is the viscous scale. The vortices fill a domain of size $L$, thus the number of degrees of freedom
$N \sim (L k_{\nu})^d \sim R_{e}^{\frac{3d}{4}}$, where $R_e$ is the Reynolds number
$R_{e} \sim  \frac{L^{\frac{4}{3}} \epsilon^{\frac{1}{3}}}{\nu}$. 

We can try to estimate the anomaly coefficient $a_{K41}$ by multiplying the number of degrees of freedom by the
anomaly coefficient $a_{vortex}$ of
a free conformally invariant vortex field theory, e.g. a CFT
of a  two-form field in $d$ dimensions  \cite{Erdmenger:1997gy}. The calculation of $a_{vortex}$
can be done using the standard $\zeta$-function method and it is on general grounds an exponentially decreasing
function of the number of space dimensions. By the Landau estimate  $N$ is an exponentially 
increasing function of  the number of space dimensions.  
It is hard at this point to draw a definite conclusion about the large $d$ behaviour of $|\bar{a}_{K41}|$ although the bounds (\ref{bd})
and (\ref{bdodd})
imply that it should grow at least linearly with $d$.

\subsection{Entanglement Entropy}

Taking a space derivative of the NS equation (\ref{NS}) and using the incompressibility condition one gets a relation between the fluid velocity
and pressure  
\begin{equation}
\nabla^2 p = -\partial_i v^j \partial_j v^i \ ,
\end{equation}
that is the pressure is non-locally related to the velocity.

The strength of intermittency is determined by a competition between the small scale cascade and the non-local 
pressure effect that couples different regions in space and tends to calm it \cite{kraichnan}. 
The choice of the branch in (\ref{branches}) and (\ref{branches2}) that is compatible with the data in three and for space
dimensions suggests that in a fixed number of space dimensions the strength of the intermittency decreases with the increase of the Euler conformal anomaly
coefficient $|a_+|$ of the effective conformal field theory of turbulence  (\ref{KPZfinal}).
 
Entanglement entropy of quantum fields is a valuable tool to quantify the entanglement between degrees of freedom at different spatial regions. If we divide the space to two parts $A$ and $B$ and construct the reduced density matrix 
$\rho_A  = Tr_{B} \rho$, the entanglement entropy is the von-Neumann entropy of $\rho_A$.
The universal part of the entanglement entropy of a CFT is proportional to the conformal anomaly coefficient. For us :
\begin{equation}
S_A \sim |a_+| \log \frac{{\cal L}}{l} \ ,
\end{equation}
where ${\cal L}$ is the scale size of $A$ and $l$ is the UV cutoff, which is the viscosity scale.
For large $|a_+|$ the entanglement entropy is large  indicating a strong correlation between the different spatial
region and hence weak intermittency. For  small  $|a_+|$ the entanglement entropy is small 
 indicating a weak  correlation between the different spatial
regions and hence strong  intermittency. 
This is compatible with the predictions of the KPZ-like formula for the anomalous scaling (\ref{KPZfinal})
if we work in the branch  $0 \leq {\cal G}^2  \leq 1$, which is the branch appropriate for the data in three and four dimensions. 

Information theory is a valuable framework for the analysis of  quantum field theory properties.  The above quantitative
analysis suggests that perhaps such tools could prove useful also if applied
to the statistical theory of incompressible fluid turbulence.

\section{Two, Three and Large Number of Dimensions}

\subsection{Two-dimensional Turbulence}

In two-dimensional incompressible fluid turbulence the energy casacde is an inverse cascade, that is the energy flux
flows to large length scales, and the numerical and experimental data are compatible with Kolmogorov linear scaling.
Hence we expect that ${\cal G}(d=2) \sim 0$. 
In \cite{SLE} the isovorticity lines of two-dimensional inverse cascade turbulence have been studied numerically and
have been identified as $SLE_{\kappa}$ curves with $\kappa =6$ (for an $SLE$ review see e.g. \cite{cardy}). This result suggests that there is
an underlying two-dimensional conformal structure in inverse cascade turbulence theory. 
The central charge of this theory can be read from $\kappa$ via
\begin{equation}
c = \frac{(8-3\kappa)(\kappa-6)}{2\kappa} \ ,
\end{equation}
and using $\kappa=6$ we get $c=0$.
The two-dimensional relativistic Euler conformal anomaly coefficient $a$ is the central charge $c$ of the theory.

This can be compatible with our analysis if there is no NG dilaton in the two-dimensional inverse cascade where the IR scale
is the viscous scale, and is perhaps analogous to non-existence of NG bosons in two-dimensional relativistic field theories. 
We associate with the two-dimensional turbulent field theory $a_{dilaton}=a_{K41}=0$ and consistently 
Eq. (\ref{a}) implies that the dilaton effective action vanishes. 
There is no dressing of the K41 operators, the bound (\ref{bd}) is not satisfied and we should not use the the KPZ relation (\ref{KPZfinal}).
The intermittency parameter vanishes and we have the Kolmogorov linear scaling.
This is reminiscent of two-dimensional 
quantum gravity \cite{Knizhnik:1988ak,Distler:1988jt} in the case where the central charge of the matter $c_{matter} = -c_{ghost} =26$
and the matter CFT decouples from gravity.

\subsection{Three-dimensional Turbulence}

The value of  the conformal anomaly coefficient that is needed in order to explain the experimental and numerical data 
of three-dimensional turbulence  is  $\bar{a}_+ \sim -12.6$ and $a_+ \sim -1$ since ${\cal G}^2 \sim 0.16$ \cite{Eling:2015mxa}, and we need to choose
the $(-)$ branch.
The main contribution to the dilaton anomaly comes from the background charge (\ref{backcharge}) :  $a_{dilaton} \simeq a_{K41} \simeq \frac{a_+}{2}$.  
It is curious that we need such a large anomaly coefficient $a_{+}$ to account for the turbulence data: $|a_{+}| \sim 10^3 a_{scalar}$, 
where $a_{scalar}$ is the boundary Euler anomaly coefficient of a free scalar in three dimensions \cite{Jensen:2015swa}.
Although we cannot perform a 
precise calculation at this point,
the estimate of $a_{K41}$ in a previous subsection can account for such a large number.

\subsection{Large $d$ Turbulence}

The analysis at large $d$ depends on the asymptotics of $a_{+}$. 
The numerical data in four space dimensions  is compatible with $\bar{a}_{+} \sim -47$ and  $a_{+} \sim - 0.15$ since ${\cal G}_-^2 \sim 0.278$ \cite{Eling:2015mxa}.
As above, $|a_{+}| \sim 10^3 a_{scalar}$ where  now $a_{scalar}$ is the Euler anomaly coefficient of a free scalar in four dimensions.
We  have an indication that  $\bar{a}_{+}$ increases while $a_{+}$ decreases as we increase $d$, which
is also the behaviour of the critical values $\bar{a}_{critical}$  and $a_{critical}$.
The large $d$ behaviour of the anomalous exponents depends on the detailed limit of $\bar{a}_{+}$. If it approaches 
 $\bar{a}_{critical}$  we will have in the limit  ${\cal G}_-^2 = G_{+}^2 =1$ and $\xi_n \sim \sqrt{n}$.
 However, it can increase to infinity faster than $\bar{a}_{critical}$ and depending on the branch we 
 can get  the suggestion of \cite{Falkovich:2009mb}  that $\xi_n \rightarrow 1$  at large $d$,
 or   $\xi_n \sim n$.

\section{Discussion and Outlook}

We proposed an effective CFT description of steady state incompressible fluid turbulence at the inertial range of scales,
which
consists of a K41 mean field CFT coupled
to a NG dilaton. 
The dilaton arises from a spontaneous breaking,  due to the energy flux,  of the separate scale and time symmetries of the inviscid Navier-Stokes equations to the K41 scaling.
We proposed that  it acts as a random measure that dresses the K41 theory and introduces intermittency.
Using this framework we derived
a KPZ-type equation for the anomalous scaling  (\ref{KPZfinal}), and
related the intermittency parameter to the boundary Euler A-type conformal anomaly coefficient  in (\ref{branches}), (\ref{intermittency})
and (\ref{branches2}).
We noted that field theory entanglement entropy can be used to characterize the intermittency strength.
Finally, we considered the physical cases of two and three space dimensions and the large $d$ limit.

There are many open questions that are worth pursuing.
The most important one is the construction of the K41 effective field theory and the calculation of the 
conformal anomaly coefficient and 
the intermittency parameter. In particular, an understanding of the physical two-dimensional and three-dimensional cases and 
the non-local structure of the odd-dimensional cases in general is required. 
A careful analysis of the boundary conditions is also needed since the value of the anomaly coefficient depends on that.
Developing the OPE and  bootstrap approach  to the flux states can lead to a significant breakthrough
in understanding the turbulence field theory (see \cite{OPE}).
Studying the consistency conditions on the KPZ scaling that follow from the NS equation as suggested in \cite{Polyakov:1995mn}
is a valuable direction to follow.
There is a mixture in the constraint equations that involve the longitudinal as well as the transversal 
structure functions  \cite{Yakhot}, which 
in order to solve requires a proposal for the form of the transverse structure functions.
The KPZ scaling  (\ref{KPZfinal}) predicts that at large $n$ and finite $d$ the anomalous scalings $\xi_n \sim \sqrt{n}$.
It would be of much importance to understand what type of fluid field configurations can lead to such a scaling.

There is  a puzzle that needs to be addressed.
If we try to use KPZ formula for the local energy dissipation $\epsilon(x)$, whose  undressed K41 dimensions $\Delta_0[\epsilon(x)] = 0$,
we get $\Delta[\epsilon(x)]  = 1- \frac{1}{{\cal G}^2}$.
But in order to match to experimental data in three space dimensions  $\Delta \sim -0.1$, this means ${\cal G}^2 \sim 1$.
The data for the structure functions $S_n$ suggest  ${\cal G}^2 \sim 0.16$ \cite{Eling:2015mxa}.  It is possible that one cannot apply  the KPZ-type formula to $\epsilon(x)$, but understanding why this is the case, and in general for which operators the KPZ formula can be applied  is important.

In one space dimension one describes (compressible) fluid flows by the Burgers equation. The steady state statistics 
is characterized by the anomalous scaling exponents
$\xi_n = 1$. It would be interesting to see whether a one-dimensional CFT on a line interval provides a field theory description
of Burgers turbulence that requires an infinite intermittency parameter.
Another interesting direction to follow is the anomalous scalings of relativistic turbulence \cite{Fouxon:2009rd,Westernacher-Schneider:2015gfa} and 
the construction of an effective field theory framework to calculate them. 

The AdS/CFT correspondence is a powerful framework to analyze strongly coupled CFTs  \cite{Aharony:1999ti} and perhaps can be 
used to study our proposal for an effective field theory of turbulence.

Finally, the field theory structure that has been discussed in this work is worth studying irrespective of whether
it provides the correct description of incompressible fluid turbulence. In particular, the higher-dimensional
generalization of the Liouville action (\ref{actionQ}) and (\ref{Liouvillefcosm}) is both an interesting CFT and may be an important ingredient in studying a summation over 
manifolds as part 
of a study of higher-dimensional random geometry and gravity \cite{Tom}.

\section*{Acknowledgements}
I would like to thank I. Arav, C. Bachas, C. Eling, I. Hason, B. Keren-Zur, T. Levy, V. Mukhanov and  A. Polyakov for discussions.
This work  is supported in part by the I-CORE program of Planning and Budgeting Committee (grant number 1937/12), the US-Israel Binational Science Foundation, GIF and the ISF Center of Excellence.

\appendix*

\section{Notations and Higher-Dimensional Liouville Field Theory}

Let $M$ be an even-dimensional manifold of dimension $d$ and metric $g_{ab}$.
Under a Weyl transformation of the metric $g_{ab} \rightarrow e^{2 \sigma} g_{ab}$ the
conformally covariant GJMS operator ${\cal P}\equiv P_{\frac{d}{2}} = \Box^{\frac{d}{2}} + lower~ order$ \cite{GJMS} transforms as :
\begin{equation}
{\cal P}_{e^{2 \sigma} g} = e^{- d \sigma} {\cal P}_{g} \ , \label{P}
\end{equation}
and the ${\cal Q}$-curvature \cite{Q}, ${\cal Q} = -\frac{1}{2(d-1)}\Box^{\frac{d}{2}-1} R + ... $  as :
\begin{equation}
{\cal{Q}}_{e^{2 \sigma} g} = e^{- d \sigma}\left({\cal{Q}}_{g} + {\cal P}_{g} \sigma \right) \ .
\end{equation}
Define the higher-dimensional Liouville action by :
\begin{equation}
S(g,\tau) = \int_M \sqrt{g} d^d x \left( \tau {\cal P} \tau  + 2 \tau {\cal{Q}}_{g} \right) \ . \label{actionQ}
\end{equation}
Under  the Weyl transformation $g_{ab} \rightarrow e^{2 \sigma} g_{ab},  \tau \rightarrow  \tau - \sigma$ we have :
\begin{equation}
S(e^{2 \sigma}g, \tau -\sigma) = S(g, \tau) - S(g, \sigma) \ . \label{anomalytransformation}
\end{equation}

We can relate the ${\cal Q}$-curvature and the Euler density $E_d$ on a constant curvature conformally flat manifold 
$M$. Using :
\begin{equation}
\int_M {\cal{Q}}_{g} = \frac{(-1)^{\frac{d}{2}}}{d} (4 \pi)^{\frac{d}{2}}  (\frac{d}{2})! \chi (M) \ ,
\end{equation}
and
\begin{equation}
\int_M E_d = (4 \pi)^{\frac{d}{2}} (\frac{d}{2})! \chi (M) \ , \label{EM}
\end{equation}
where $\chi (M)$ is the Euler characteristic of $M$ and $E_d$ is the Euler density,
we obtain on a constant curvature conformally flat manifold the relation :
\begin{equation}
E_d  = (-1)^{\frac{d}{2}}  d {\cal Q}_{g}  \ . \label{EQ}
\end{equation}
Since $\chi(S^d)=2$ we get from (\ref{EM}) 
\begin{equation}
\int_{S^d} E_d = \Omega_d d! \ , \label{normalization}
\end{equation}
where 
\begin{equation}
\Omega_d = \frac{2 (\pi) {\frac{d+1}{2}}}{\Gamma[\frac{d+1}{2}]}  =  \frac{2 (4 \pi)^{\frac{d}{2}} (\frac{d}{2})!} {d!}  , \label{O}
\end{equation}
is the surface volume of the $d$-sphere.

Using (\ref{actionQ}) and (\ref{EQ}) we define :
\begin{equation}
S_{Liouville} (g,\tau)  = \int_{M} \sqrt{g} d^d x  \left(\frac{d}{2} \tau\Box^{\frac{d}{2}}\tau   + (-1)^{\frac{d}{2}}  \tau E_d \right)  \ , \label{Liouvillef}
\end{equation}
Using (\ref{anomalytransformation}) we have under Weyl transformation to order $O(\sigma^2)$ :
\begin{equation}
\delta_{\sigma} S_{Liouville} (g,\tau) =  \int_{M} \sqrt{g} d^d x  \sigma T^{\mu}_{\mu}  =  \int_{M} \sqrt{g} d^d x  \sigma\left( - (-1)^{\frac{d}{2}} E_d \right) \ .
\label{ds}
\end{equation}
Hence, the anomaly action $a S_{Liouville}$
where $a$ is the anomaly coefficient gives :
\begin{equation}
\langle T^{\mu}_{\mu} \rangle = - (-1)^{\frac{d}{2}} a E_d  \ . \label{wz}
\end{equation}
Note that in $d > 2$ there are addition terms in the WZ anomaly action compared to (\ref{Liouvillef}), which in our work were 
surpressed since we worked in the inertial range of scales $r \ll L$, where $L$ is the size of the ball and the boundary sphere.
(\ref{wz}) follows from
solving the Wess-Zumino consistency conditions and is independent of whether the CFT is unitary.

The Euler characteristic of an even-dimensional  manifold of dimension $d$ with a boundary reads :
\begin{equation}
\chi (M) =   \frac{2}{\Omega_d d!} \left(\int_M E_d  -  \int_{\partial M} q_d \right) \ , \label{Eback}
\end{equation}
where $\partial M$ is the boundary of $M$  and
$q_d$ is a boundary Euler class term constructed from the boundary intrinsic and extrinsic curvatures.
When $M=B_d$  is the $d$-dimensional ball in  Euclidean space, whose boundary is the $d$-sphere  $\partial M = S^{d-1}$,  the bulk Euler term $E_d$ is zero in (\ref{Eback}).
The ball $B_d$ is contractible, hence its Euler characteristic is that of the point $\chi(B_d) =1$,  
\begin{equation}
\chi(S^{d-1}) = \left(1 + (-1)^{d-1}\right) \chi(B_d) \ ,
\end{equation}
which  fixes the normalization  of $q_d$ as :
\begin{equation}
 - \int_{\partial B_d}  q_d  = \frac{1}{2} \Omega_d d! = \alpha_d \  . \label{chiBa}
\end{equation}
The Liouville action (\ref{Liouvillef}) and its conformal transformation (\ref{ds}) in the presence of the boundary
are defined in the same way.

In general, one can also define a boundary 
 ${\cal P}^{b}\equiv P_{\frac{d-1}{2}}$ that  transforms as  \cite{odd} :
\begin{equation}
{\cal P}^{b}_{e^{2 \sigma} g} = e^{- (d-1) \sigma} {\cal P}^b_{g} \ , 
\end{equation}
and a boundary ${\cal Q}$-curvature such that :
\begin{equation}
{\cal{Q}}^{b}_{e^{2 \sigma} g} = e^{- (d-1) \sigma}\left({\cal{Q}}^{b}_{g} + {\cal P}^{b}_{g} \right) \ .
\end{equation}

When $M$ is an odd-dimensional manifold of dimension $d$ one can define a pseudo-differential 
conformally covariant GJMS operator $P_{\frac{d}{2}} = \Box^{\frac{d}{2}} + lower~ order$, which transforms under
Weyl transformation as in (\ref{P}) (see e.g. \cite{fractional}). It is plausible that one can define 
the analog of the ${\cal Q}$-curvature and the action (\ref{actionQ}) in this case. We are not aware, however, of such a construction.

Finally, note that we can add a "cosmological constant" term to the Liouville action  (\ref{actionQ}) of the form :
\begin{equation}
S_{cosmo} (g,\tau)  = \int_{M} \sqrt{g} d^d x~  e^{d \tau}  \ , \label{Liouvillefcosm}
\end{equation}
which we did not use in this work, but will be valuable in the study of the higher-dimensional Liouville theory.

\end{document}